\documentclass[useAMS,usenatbib]{mn2e}

\usepackage{epsf}

\newcommand{\bmth}[1]{\mbox{\boldmath${#1}$}}

\title[Oscillations of rotating bodies: A self-adjoint formalism
applied to dynamic tides and tidal capture]{Oscillations of rotating
bodies: 
A self-adjoint  formalism applied to dynamic tides and tidal capture}
\author[J. C. B. Papaloizou and P. B. Ivanov]{J.C.B.Papaloizou$^{1}$\thanks{E-mail:
J.C.B.Papaloizou@damtp.cam.ac.uk (JCBP);  P.Ivanov@damtp.cam.ac.uk (PBI)} and P. B. Ivanov
$^{1,2}$\footnotemark[1]\\
$^{1}$Department of Applied Mathematics and Theoretical Physics,
University of Cambridge,\\
Centre for Mathematical Sciences, 
Wilberforce Road, Cambridge, CB3 0WA, UK \\
$^{2}$Astro Space Centre, P. N. Lebedev Physical Institute,
 4/32 Profsoyuznaya Street,
 Moscow, 117810, Russia}
\begin{document}

\date{Accepted Received ; in original form }

\pagerange{\pageref{firstpage}--\pageref{lastpage}} \pubyear{2002}

\maketitle

\label{firstpage}

\begin{abstract}
We consider the excitation of the  inertial modes of 
a uniformly rotating fully convective body  
due to a close encounter with another object. This
could lead to a tidal capture  or orbital circularisation
depending on whether the initial orbit is unbound or highly
eccentric.
We develop a  general self-adjoint formalism for the response
problem and thus solve it taking into account
the inertial modes with  $m=2$ for a full polytrope with $n=1.5.$
We are accordingly 
able to show in this case  that the excitation of inertial modes dominates
the response for large impact parameters and 
thus  cannot be neglected in calculations of tidal
energy and angular momentum exchange 
 or orbital circularisation from large eccentricity.
\end{abstract}

\vspace{-1cm}
\section{Introduction}
The process of tidal capture of a body  into a bound orbit
through the excitation
of oscillatory internal modes  is thought to be
of general importance in astrophysics.
It is likely to play a role in binary formation 
in globular clusters (eg. Press \& Teukolsky, 1977,  hereafter PT
and references therein) as well as
in tidal interactions of stars within galactic centres. When there are
repeated close encounters as for a highly eccentric orbit,
circularisation may result, a process of potential
importance for extrasolar planets (eg. Ivanov \& Papaloizou 2004, hereafter IP).

Until now, only  the oscillation modes
associated with spherical non rotating stars
have been considered with possibly rotation
being treated as a perturbation (eg. PT,IP).  But angular momentum is transfered
in tidal encounters and when, as is often the case,
the internal inertia  is much less than that of the orbit,
the tidally excited object would be expected
to rotate at a balanced rate or undergo pseudo synchronisation.
Then the rotation frequency is matched to the most important
ones in a Fourier decomposition of  the tidal forcing.
It is  then natural to   expect that inertial modes controlled by the rotation
(see eg. Papaloizou \& Pringle 1981, hereafter PP)
become important for the tidal response. This is especially
the case for a barotropic configuration 
\footnote{For a barotropic fluid the pressure $P$ has the
same functional dependence on the density $\rho$: $P=P(\rho)$, 
for the unperturbed configuration and during perturbation.}
, with no stratification
and hence no  low frequency $g$ modes. 
We would expect inertial modes, which have  periods
that scale with the rotation frequency, to be particularly
important for large periastron distances as the 
fundamental spherical modes suffer increasing frequency mismatch
as this distance increases.

It is the purpose of this paper to calculate the 
response of a rotating body taking the inertial modes into
account. In particular we focus on modes with $m=2$
which dominate the angular momentum exchange and a full polytrope
with index $ n= 1.5.$
We show that the inertial modes indeed dominate the response
at larger periastron distances of interest in many cases.

In section 2 we give the basic equations governing the
linear response of a uniformly rotating barotropic star to
the tidal
forcing due to a close encounter. In section 3 we give a self-adjoint
representation of the linear problem and discuss how it can
be used to provide a formal solution in terms of a spectral
decomposition as can be done for a non rotating star.
In section 4 we  develop expressions 
for the energy and angular momentum exchange. In section 5 
we give results for a polytrope with index $n=1.5$,
showing that the $m=2$  inertial  wave  response is contained
within just a few global modes and that inertial modes dominate the tidal
response for large periastron passages. Accordingly these
must be considered when discussing tidal circularisation from 
large eccentricity for objects such as extrasolar planets.
\vspace{-0.7cm}
\section{Basic equations}
We assume that the star is rotating with uniform angular velocity
$\bmth {\Omega }$ directed perpendicular to the orbital plane. 
In this case the
hydrodynamic equations for the perturbed quantities 
take the simplest form in the rotating
frame. For a convective star, approximated as barotropic, we have
\begin{equation}
{D^{2} {\bmth{ \xi}} \over Dt^{2}}+2{\bmth {\Omega}}\times {D \bmth
{\xi}\over Dt}=-\nabla W. \label{eq 1} 
\end{equation}
Here ${\bmth {\xi}}$ is the Lagrangian displacement vector, and 
\begin{equation}
W=c_{s}^{2}\rho^{'}/\rho+\Psi^{int}+\Psi^{ext},  \label{eq 2}
\end{equation}
where $\rho $ is the density, $\rho^{'} $ is the
density perturbation, $c_{s}$ is the 
adiabatic sound speed, $\Psi^{int}$ is the  stellar  gravitational 
potential  arising from the  perturbations and  $\Psi^{ext}$ is the
 external forcing  tidal
potential. Note that the centrifugal term is absent in equation  $(\ref{eq 1})$
being formally incorporated into the potential governing
the static equilibrium of the unperturbed star.
The convective derivative ${D\over Dt} \equiv {\partial \over \partial t}$
as there is no unperturbed motion in the rotating frame.

\noindent The linearised continuity and Poisson  equations are    
\begin{equation}
\rho^{'}=-\nabla (\rho {\bmth {\xi}}) \quad {\rm and} \quad \Delta \Psi^{int}=4\pi
G\rho^{'}.  \label{eq 3}
\end{equation}

\noindent We use  a
cylindrical coordinate system $(\varpi, \phi, z)$ with origin at the centre
of the star  and assume
that the tidal potential and the resulting
perturbation to  some  quantity, say $Q$,
is represented in terms of a  Fourier  transform over the angle 
$\phi$ and time $t$.
\noindent  Thus
$Q = \sum_{m}\left(  \exp({im\phi})\int^{+\infty}_{-\infty}d\sigma \tilde
Q_{m}\exp({-i\sigma t}) + cc \right )$,
where the sum is over $m=0$  and $2$ and $cc$ denotes
the complex conjugate. The reality of $Q$
implies that the   Fourier transform, indicated by tilde  satisfies 
$\tilde Q_{m}(\sigma) = \tilde
Q_{-m}^*(-\sigma).$

\noindent To solve the forcing problem we obtain equations for the
Fourier transforms of the  perturbations using  
 $(\ref{eq 1}) - (\ref{eq 3}).$   
We can then express $\tilde {\bmth {\xi}}_{m}$ in terms of 
$\tilde W_{m}$ with help of
$(\ref{eq 1})$, the density perturbation in terms of $\tilde \Psi^{ext}_{m}$,
$\tilde \Psi^{int}_{m}$ and $\tilde W_{m}$ 
with help of (\ref{eq 2}), and substitute the 
results into the continuity equation  (\ref{eq 3}) to get
\begin{equation} 
\sigma^{2}  {\bmth{A}} \tilde W_{m} -\sigma {\bmth{B}}
\tilde W_{m} -{\bmth{C}}\tilde W_{m} =
\sigma^{2}d{\rho \over c_{s}^{2}}
(\tilde \Psi^{ext}_{m}+\tilde \Psi^{int}_{m}-\tilde W_{m}),  \label{eq 4}
\end{equation}
where $d=4\Omega^{2}-\sigma^{2}$, and
\begin{equation} 
{\bmth {A}}=-{1\over \varpi}{\partial \over \partial \varpi}\left (\varpi \rho
{\partial\over \partial \varpi}\right )-{\partial \over \partial z}\left (\rho
{\partial \over \partial z}\right )+
{m^{2} \rho \over \varpi^{2}},  \label{eq 5}
\end{equation}
\begin{equation} 
{\bmth {B}} =-{2m\Omega \over \varpi}{\partial \rho \over \partial \varpi}, 
\quad {\bmth {C}}=-4\Omega^{2}{\partial \over \partial z}\left(\rho {\partial
\over \partial z}\right).  \label{eq 6}
\end{equation}
It is very important to note that the operators ${\bmth {A }}$, 
${\bmth {B}}$ and ${\bmth {C}}$ are self-adjoint  when the inner product 
\begin{equation}
(W_{1}|W_{2})=\int_{V} dz \varpi d\varpi W^{*}_{1}W_{2},  \label{eq 7}
\end{equation} 
with $V$ denoting the volume of the star. Also when $W$ is
not a constant
${\bmth {A}}$ and ${\bmth {B}}$ are positive 
definite and  ${\bmth {C}}$ is non negative definite. 
Equations (\ref{eq 2}), (\ref{eq 4}) and the Poisson equation 
(\ref{eq 3}) form a complete set.

\noindent When $\tilde \Psi^{ext}_{m} = 0,$ equation (\ref{eq 4})
describes  the free oscillations of a rotating star. These may be classified as
relatively high frequency $f$ and $p$ modes and inertial modes with 
eigen frequencies $\sigma \sim \Omega$. The $f$ and $p$ modes  exist
in non rotating stars and can be
treated in a framework of  perturbation theory  (eg. IP).
Here we 
focus on the response of inertial waves to  tidal
forcing, for slow rotation. We seek a
solution  as a series expansion  in the small parameters 
$\sigma /\Omega_{*}$ and $\Omega/\Omega_{*}$, where $\Omega_{*}=\sqrt
{(GM_{*}/R^{3}_{*})}$, with 
 $M_{*}$  and $R_{*}$  being the mass and the radius of the
star, respectively. The small parameters 
will be assumed to be of the same order. 

\noindent To zeroth order (when $\sigma \rightarrow 0$)  the tidal potential
induces a quasi-static bulge in which 
tidal forces are balanced by pressure and self gravity
and $\tilde \rho^{'}_{m} \rightarrow \tilde \rho_{m,st}$ with 
${\tilde \Psi^{int}_{m}} \rightarrow {\tilde \Psi^{int}_{m,st}}$.
From equation (\ref{eq 4}) we conclude that $W$ is 
smaller than the potential perturbations by a factor
of order $\sigma^2/\Omega^{2}_{*}$ (see PP for details).
Accordingly, setting $W=0$
in equation  (\ref{eq 2}) and using equation (\ref{eq 3}) we  obtain
\begin{equation}
\tilde \rho_{m,st}=-{\rho\over c^{2}_{s}}(\tilde
\Psi^{int}_{m,st}+\tilde \Psi^{ext}), 
\quad \Delta \Psi^{int}_{m,st}=4\pi G\tilde \rho_{m,st}.  \label{eq 8}
\end{equation}
In the  lowest order 
 approximation we find $\tilde W_{m}$ using equation  (\ref{eq 4})
 neglecting the  higher  order term proportional to
 $\rho \tilde W_{m}/c_{s}^{2}$  
 on the right hand
side. We  thus have
\begin{equation}
\sigma^{2} {\bmth {A}} \tilde W_{m} -\sigma {\bmth {B}}
 \tilde W_{m} 
-{\bmth {C}} \tilde W_{m}=S,  \label{eq 9}
\end{equation}  
where the source term 
\begin{equation}
S=-\sigma^{2}d\tilde \rho_{m, st}=\sigma^{2}d{\rho \over c_{s}^{2}}
(\tilde \Psi^{ext}_{m}+\tilde \Psi^{int}_{m,st})  \label{eq 10}
\end{equation}
is completely determined
by the external forcing via equation (\ref{eq 8}).
Finally,  the perturbation of the
internal gravitational potential 
associated with $\tilde W_{m}$,  
$\delta \tilde \Psi_{m}=\tilde \Psi^{int}_{m}-\tilde \Psi^{int}_{m,st}$   
follows from (\ref{eq 2}),  (\ref{eq 3}) and (\ref{eq 8})
\begin{equation}
\Delta \delta \tilde \Psi_{m}=
{4\pi G \rho \over c_{s}^{2}}(\tilde W_{m}-\delta \tilde \Psi_{m}).
\label{eq 11}
\end{equation}
\vspace{-0.7cm}
\section{Reduction  of the problem to  standard
 self-adjoint form and formal solution}
Equation (\ref{eq 9}) is of a general type that governs the linear response
of rotating bodies in a context that is much wider
than the specific one considered here ( see Lynden Bell \& Ostriker 1967).
One recurring difficulty has been that the  way in which the frequency
$\sigma$ occurs does not readily allow application of the general
spectral theory of self-adjoint operators to the response calculation
even though     
${\bmth {A}}$, 
${\bmth {B}}$ and ${\bmth {C}}$ are themselves 
self-adjoint (eg. Dyson \& Schutz 1979).
We here provide a formulation in which  the problem can be 
dealt with in the same way as a standard self-adjoint one such as occurs
for a non rotating star.

\noindent We begin by remarking that the self-adjoint and non negative
 character of ${\bmth {A}}$, ${\bmth{B}}$ and ${\bmth{C}}$ allows
the spectral theorem to be used to specify
their square roots, eg. ${\bmth {A}}^{1/2}$,
defined by condition ${\bmth {A}}={\bmth {A}}^{1/2}
{\bmth {A}}^{1/2}$. The positiveness of ${\bmth {A}}$ allows
definition of the inverse of ${\bmth {A}}^{1/2}$, ${\bmth {A}}^{-1/2}$.
Technically, this is done through using  the spectral decomposition 
\begin{equation}
{\bmth {A}}W= \sum_{j}  \lambda_j \phi_j \left< W|\phi_j\right>
\quad 
F({\bmth {A}})W= \sum_{j}  F(\lambda_j) \phi_j \left< W|\phi_j\right>,
\label{spectra}\end{equation}
..etc., where $\phi_{j}$ and $\lambda_{j}$ are the eigenfunctions and
the eigenvalues of the corresponding operators and the last equation 
defines a general function of the operator ${\bmth {A}}$. The sum
may formally include integration with an appropriate measure in the case of
a continuous spectrum of the corresponding operators. 

\noindent Now let us consider a generalised  two dimensional 
vector  $\vec Z$ with components 
such that $ \vec Z  =
(Z_1 ,{\bmth {C}}^{1/2}  
\tilde W_{m})$ and the straightforward generalisation of the  inner product
 given by equation (\ref{eq 7}). 
It is easy to see that equation (\ref{eq 9}) can be  derived from  the 
matrix form
\begin{equation}
\sigma \vec Z ={\bmth {H}} \vec Z +\vec S,  \label{eq 12} 
\end{equation}
where
\begin{equation}
{\bmth {H}}=  
\left( \begin{array}{cc} {\bmth {A}}^{-1/2}{\bmth {B}}
{\bmth {A}}^{-1/2} & 
{\bmth {A}}^{-1/2} {\bmth {C}}^{1/2} \\
{\bmth {C}}^{1/2} {\bmth {A}}^{-1/2} & 0 \end{array}\right), \label{eq 13}
\end{equation} 
and the source vector $\vec S$ has the components
$({\bmth {A}}^{-1/2}S, 0).$
It then  follows from (\ref{eq 12})  that we may choose  
$Z_1  =  \sigma {\bmth {A}}^{1/2}  \tilde W_{m}.$ Since the off
diagonal elements in the matrix (\ref {eq 13}) are adjoint of each
other and the diagonal elements are self adjoint, 
that the operator  ${\bmth{ H}}$ is self-adjoint is manifest. Now we can use 
the spectral  theory of self-adjoint operators  to look for a
solution to  (\ref{eq 9}) in
the form $\vec Z=\sum_{k} \alpha_{k} \vec Z_{k}$, where 
 $\vec Z_{k}$ are the real eigenfunctions of ${\bmth{ H}}$ which satisfy
\begin{equation}  
\sigma_{k} \vec Z_{k} ={\bmth{ H}} \vec Z_{k},  \label{eq 14} \end{equation}
 the associated necessarily real 
eigen frequencies being  $\sigma_k.$ 

\noindent Substituting  (\ref{eq 14}) into  (\ref{eq 12}) we obtain
\begin{equation}  
\alpha_{k} ={<\vec Z_{k}| \vec S>\over <\vec Z_{k}|\vec
Z_{k}>(\sigma+i\nu-\sigma_{k})},  \label{eq 15}
\end{equation}
where, following the Landau prescription,  we have
added an infinitesimal  imaginary part  $\nu >  0$ to the  frequency
$\sigma.$  The inner product induced by
${\bmth{ H}}$   turns out to be
$<\vec Z_{k}| \vec Z_{l}>=\sigma_{k}\sigma_{l}(W_{k}| {\bmth {A}}
W_{l})+(W_{k}|{\bmth {C}} W_{l})$, where $W_{k}={\bmth
{C}}^{-1/2}Z_{2}^{k}=\sigma_{k}^{-1}{\bmth A}^{-1/2}Z_{1}^{k}$.
Using  (\ref{eq 10}) and  (\ref{eq 15})
we explicitly obtain
\begin{equation}
\tilde W_{m}=
\sigma d\sum_{k} {\sigma_{k}^{2} (W_{k}| \tilde \rho_{m,st}) 
\over N_{k}(\sigma_{k}-i\nu-\sigma)}W_{k},  \label{eq 16}  
\end{equation}
where $N_{k}=\sigma^{2}_{k}(W_{k}| {\bmth {A}}
W_{k})+(W_{k}|{\bmth {C}} W_{k})$ is the norm.

\vspace{-0.5cm}
\section{Transfer of energy and angular momentum from orbit to inertial
waves}

The time derivative $\dot E$ of the wave energy $E$ associated with the
rotating frame and the time derivative  $\dot L$ of the wave 
angular momentum  $L$ are given by usual expressions
\begin{equation}
\dot E=-\int d\phi \left (\dot \rho^{'}| \Psi_{ext}\right ), 
\quad \dot L=-\int d\phi \left (\rho^{'}| {\partial \Psi_{ext} \over
\partial \phi}\right ).  \label{eq 17}
\end{equation}
The wave energy associated with the inertial frame, $E_{I}$ is
given in terms of $E$ and $L$ (eg. Friedman $\&$ Schutz 1978,
hereafter FS) through
\begin{equation}
E_{I}=E+\Omega L.  \label{eq 18}
\end{equation}

Expressing $\rho^{'}$ with help of  (\ref{eq 2}) in terms of $W$ and discarding
the total derivatives which cannot contribute to the total
energy transfer $\Delta E=\int^{+\infty}_{-\infty}dt \dot E$, we  obtain
\begin{equation}
\Delta E=- \int^{+\infty}_{-\infty}dt
\int^{2\pi}_{0}d\phi\left({\rho \over c^{2}_{s}}(\dot W - \delta \dot \Psi)|
\Psi^{ext}\right).  \label{eq 19}
\end{equation}
Contributions $\Delta E_{m}$  to $\Delta E$  arising from  different   values 
of $m$ may be calculated separately.   
Substituting the corresponding
Fourier transforms 
into  (\ref{eq 19}), and integrating  
over  time, we obtain
\begin{equation}
\Delta E_{m}=4\pi^{2}i\int d\sigma \sigma \lbrace
(\tilde \Psi^{ext}_{m}|\delta \tilde \rho_{m})-
(\delta \tilde \rho_{m}|\tilde \Psi^{ext}_{m})\rbrace,  \label{eq 20}
\end{equation}
where $\delta \tilde \rho_{m}={\rho \over c^{2}_{s}}(\tilde W_{m}-\delta
\tilde \Psi_{m})$. Note that for the problem on hand
$\Psi^{ext}_{m}$ and $\Psi^{int}_{m,st}$ are real (see below).
Now we substitute the spectral decomposition  
$\delta \tilde \Psi_{m}=\sum_{k}\alpha_{k}\delta \Psi_{k}$, where 
$\Delta \delta \Psi_{k}={4\pi G\rho \over c^{2}_{s}}(W_{k}-\delta
\Psi_{k})$, 
together with  (\ref{eq 16})
into (\ref{eq 20}). The singularities in the  integral  over $\sigma$
corresponding to $\sigma=\sigma_{k}$ and determining all non-zero
contributions to the integral
are dealt with using the Landau
prescription.
Then the  energy transfer is a sum
of contributions  associated with each mode, being given by
\begin{equation}
\Delta E_{m}=-8\pi^{3}\sum_{k} {\sigma_{k}^{4}(4\Omega^{2}-\sigma^{2}_{k})\over
N_{k}}(W_{k}| \tilde \rho_{m,st})(\delta \rho_{m,k}| \tilde
\Psi^{ext}_{m}),  \label{eq 22}
\end{equation}
where $\delta \rho_{m,k}={\rho \over c_{s}^{2}}(W_{k}-\delta
\Psi_{k})$, and both integrals are  evaluated for $\sigma=\sigma_{k}$.  

\noindent Using the symmetry properties of the operator 
$\Delta +{4\pi G\rho\over c_{s}^{2}}$
one can readily see that the integrals are, in fact, the negatives of each
other,   thus we finally obtain
\begin{equation}
\Delta E_{m}=8\pi^{3}\sum_{k} {\sigma_{k}^{4}(4\Omega^{2}-\sigma^{2}_{k})\over
N_{k}}A^{2}_{k},  \label{eq 23}
\end{equation}
where $A_{k}=|(\tilde \rho_{m,st}| W_{k})|$.  

\noindent The expression for the angular momentum transfer  follows immediately  from
(\ref{eq 23}) and the general relation between wave energy and angular momentum
corresponding to a mode with   particular value of $m$:
$E_{m}/L_{m}=\sigma/m$ (eg. FS). We have $\Delta L_{0}=0$ and 
\begin{equation}
\Delta L_{2}=16\pi^{3}\sum_{k} {\sigma_{k}^{3}(4\Omega^{2}-\sigma^{2}_{k})\over
N_{k}}A^{2}_{k}.  \label{eq 24}
\end{equation}

The expressions  (\ref{eq 23}),  (\ref{eq 24}) can be  
evaluated  when the star is in a
parabolic orbit, being the limit of a highly eccentric
orbit, using the
expression for $\tilde \Psi^{ext}$  given in 
 the quadrupole approximation by PT. From this we obtain
\begin{equation}
\tilde \Psi^{ext}_{m}=\sqrt{{5\over 2\pi^{3}}{(2-|m|)!\over (2+|m|)!}}            
{B_{|m|}\over
(1+q)}I_{2,-m}(y) \Omega_{p}r^{2}P^{|m|}_{2}(\cos \theta ),  \label{eq 25} 
\end{equation} 
where $B_{2}=\sqrt{{3\pi\over 10}}$, $B_{0}=-{1\over 2}\sqrt{{\pi\over 5}}$.
$q$ is the ratio of $M_{*}$
to the mass of the star exerting the tides, $M$: $q=M_{*}/M$. 
$\Omega_{p}=\sqrt{{GM(1+q)\over R^{3}_{p}}}$ is a typical frequency of
periastron passage and $R_{p}$ is the periastron distance. The
functions $I_{2,-m}(y)$
specify the dependence of $\tilde \Psi^{ext}_{m}$ on $\sigma$, they are
described by PT. y=$\bar \Omega (\bar \sigma +m)$,
where we use dimensionless frequencies $\bar \sigma =\sigma/ \Omega$ 
and $\bar\Omega =\Omega/\Omega_{p}$. 
$P_{l}^{m}(x)$ is the associated Legendre function, 
and $(r,\theta)$ are the spherical coordinates.  

\noindent Now we express the density perturbation entering in  (\ref{eq 23}),
(\ref{eq 24}) as 
$\tilde \rho_{m,st}=-{\rho \over c_{s}^{2}}F\tilde \Psi^{ext}_{m}.$
Here we introduce a factor $F$ to  account for self-gravity of the
star that can be  obtained from the solution of (\ref{eq 8})
\footnote{$F=1$ in the frequently used~ 
Cowling approximation.}~. We substitute this expression and  
(\ref{eq 25}) in (\ref{eq 23}), (\ref{eq 24}) and express all quantities
 in  natural 
units. These are such that the spatial coordinates, density and 
sound speed are expressed in units of $R_{*}$, 
the averaged density 
${\bar \rho}={3M_{*}\over 4\pi R_{*}^{3}}$, and $\sqrt {GM_{*}\over
R_{*}}$, respectively. We then  have
\begin{equation} 
\Delta E_{m}={C_{m}\over (1+q)^{2}}\bar {\Omega}^{4}\sum_{k} \lbrace 
\bar \sigma_{k}^{4}(4-\bar \sigma_{k}^{2})Q_{k}^{2}
I^{2}_{2,-m}(y) \rbrace {E_{*}\over \eta^{6}},  \label{eq 26}
\end{equation}
\begin{equation}
\Delta L_{2}={2C_{2}\over (1+q)^{2}}\bar {\Omega}^{3}\sum_{k} \lbrace 
\bar \sigma_{k}^{3}(4-\bar \sigma_{k}^{2})Q_{k}^{2}
I^{2}_{2,-m}(y) \rbrace {L_{*}\over \eta^{5}},  \label{eq 27}
\end{equation} 
where $E_{*}={GM_{*}\over R_{*}}$, $L_{*}=M_{*}\sqrt{GM_{*}R_{*}}$, and
$\eta ={\Omega_{*}/ \Omega_{p}}=
\sqrt{M_{*}R_{p}^{3}\over (1+q)MR_{*}^{3}}$, $C_{2}={3\over 16}$ and 
$C_{0}=3/4$. The overlap integrals 
$Q_{k}$ have the form:
\begin{equation}
Q_{k}=({\rho\over c_{s}^{2}}Fr^{2}P^{m}_{2}|W_{k})/\sqrt{\bar N_{k}}, 
\label{eq 27a}
\end{equation}
where $\bar N_{k}=N_{k}/\Omega^{2}$.

\vspace{-0.7cm} 
\section{Numerical Calculations}

\begin{figure}
\begin{center}
\vspace{8cm}\includegraphics{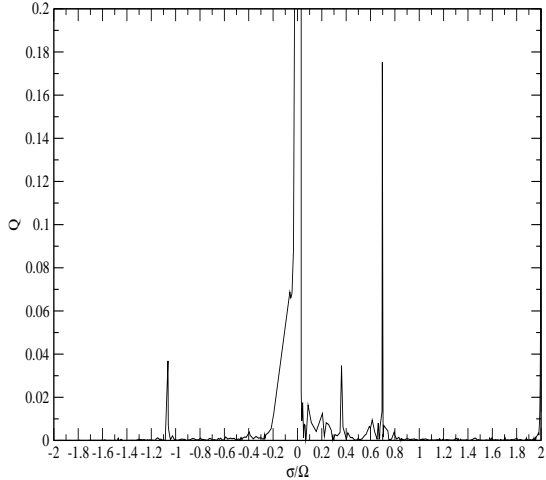}
\end{center}
\vspace{-1cm}
\caption{The dependence of $Q_{k}$ on the values of $\sigma_{k}$.}
\label{theor}
\vspace{-0.5cm}
\end{figure}

The eigen frequencies $\bar \sigma_{k}$ and the integrals $Q_{k}$ 
in equations  (\ref{eq 26}-\ref{eq 27}) have to be
calculated numerically, for a given model of the star. Therefore, 
equation  (\ref{eq 14}) describing free oscillations of the star
must be solved. For our
numerical work we model the star as $n=1.5$ polytrope and use the
method of solution of (\ref{eq 14}) proposed by PP
\footnote{In order to check the method we also calculate the eigen spectrum 
 of an $n=1$ polytrope. Our results are in a good agreement
with  results obtained by Lockitch $\&$
Friedman (1999)
 and Dintrans $\&$ Ouyed (2001) who used different methods.}. Namely, we reduce
(\ref{eq 14}) to 
an ordinary matrix equation using 
an appropriate set of trial functions, and solve 
the corresponding eigenvalue problem. The details of our approach
will be given in a separate publication. Here we note
that we use different
numbers $N_{t}$ of the trial functions to study convergence of our
numerical scheme with the largest number being equal to 
$N^{max}_{t}=225$.  
Also we adopt a spherical model of the star assuming that the 
distortion of the stellar structure due to rotation does not influence
significantly our results. Since the modes corresponding to $m=2$
fully determine the transfer of angular momentum in the quadrupole 
approximation and apparently
dominate the energy transfer, which can only be increased if $m=0$  modes are
included,
we consider only them.

\begin {figure}
\begin{center}
\vspace{8cm} \includegraphics{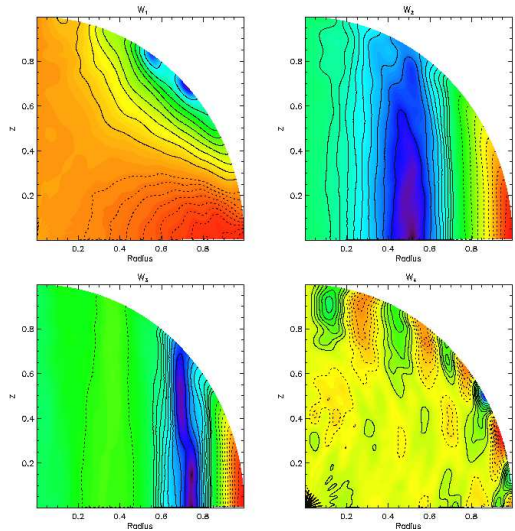}
\end{center}
\vspace{-0.4cm}
\caption{The distribution of $\tilde W_{k}$ over the  star. 
$\tilde W_{k}=0$ at the origin. Contour lines of different style
correspond to values of $\tilde W_{k}$ of opposite signs.  
The upper left 
(right) plot corresponds to $\sigma_{1}=-1.063$ (
$\sigma_{2}=0.697$). 
The lower left (right) plot corresponds to  
$\sigma_{3}=0.362$ ($\sigma_{4}=0.703$).}
\label{theor1}
\vspace{-0.4cm}
%\end{center}
\end{figure}

\begin{figure}
\begin{center}
\vspace{8cm}\includegraphics{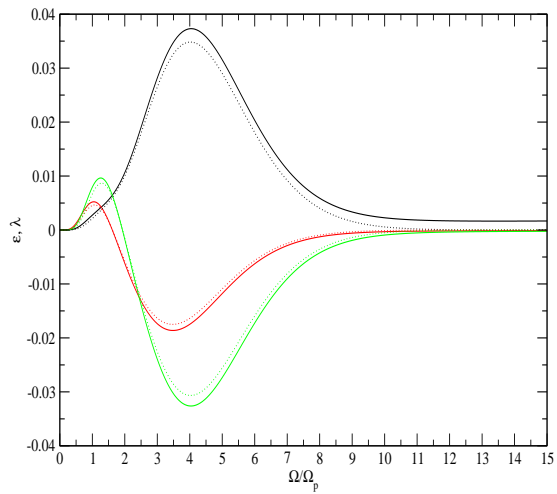}
\end{center}
\vspace{-1cm}
\caption{The dependence of the dimensionless quantities
$\epsilon$, $\epsilon_{I}$ and
$\lambda $ on $\bar \Omega$ (see (\ref{eq 28}) and the text for definitions).}
\label{theor2}
\vspace{-0.65cm}
\end{figure}

\begin{figure}
\begin{center}
\vspace{8cm}\includegraphics{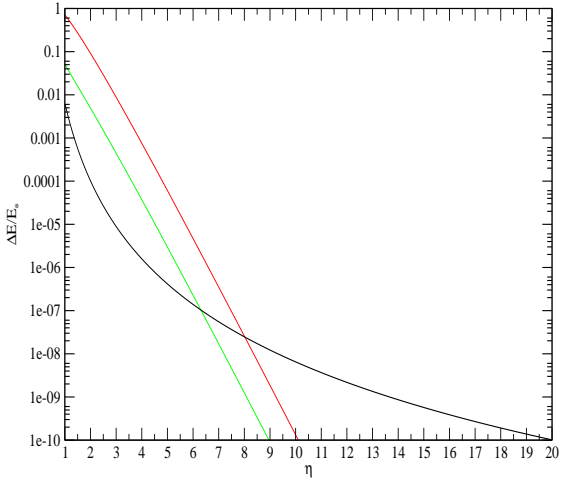}
\end{center}
\vspace{-1.2cm}
\caption{The dependence of the energy transfer on 
 $\eta$. The black curve (red and green)
give the contribution of the 
inertial waves (fundamental mode). 
We assume that $q \ll 1$ and $\bar \Omega =\bar \Omega_{crit}\approx
1.6$.}
\vspace{-0.6cm}
\label{theor3}
\end{figure}

The absolute values of the eigen frequencies $\sigma_{k}$ are
always  smaller than or equal to $2\Omega$.
For complete polytropes it is expected that they ultimately
yield an everywhere dense discrete spectrum (PP).
 In our numerical approach 
their number is equal to
$2N_{t}$, and therefore the number of the terms in the series in (\ref{eq 26})
and (\ref{eq 27}) depends on $N_{t}$. However, only the eigenfunctions
with some definite values of $\sigma_{k}$ give a
significant contribution to the series. These terms correspond to
global eigen modes with a large scale distribution of $\tilde W_{k}$ 
over the star. Therefore, the overlap integrals corresponding to
the global modes are not averaged to small values after integration
over the volume of the star. The number of the global modes does not
depend on $N_{t}$. In figure 1 we show the overlap
integrals as functions of the eigen frequencies. 
There is a sharp rise of $Q_{k}$ near $\sigma_{k}=0$ and very close
to $\sigma_{k}=2$, and several peaks corresponding to different values 
of $\sigma_{k}$. The rise at $\sigma_{k} \approx 0$, $2$ is due to
numerical inaccuracy of our method, but the corresponding modes 
practically do not influence  
our results and therefore the numerical 
inaccuracy is not significant for our purposes. 
Note that the feature at $\sigma_{k} \approx 0$ can be removed by
an appropriate regularisation of  ${\bmth{C}}$.

Three peaks with 
$\sigma_{1}=-1.063$ and $Q_{1}=0.037$, $\sigma_{2}=0.697$ and 
$Q_{2}=0.175$, $\sigma_{3}=0.362$ and $Q_{3}=0.034$ correspond to the
global modes. The spatial distribution of the corresponding eigenfunctions
is shown in figure 2. For comparison we
also show a non global mode with $\sigma_{4}=0.703$ and a very small
value of $Q_{4}$. It is clear from this figure that the global modes have
a large scale distribution of $\tilde W_{k}$.

It follows from (\ref{eq 26}) and (\ref{eq 27}) that 
for a given model of the star, the quantities
\begin{equation}
\epsilon_{m} =(1+q)^{2}\eta^{6}(\Delta E_{m}/E_{*}),
\quad \lambda =(1+q)^{2}\eta^{5}(\Delta L_{2}/L_{*}),
\label{eq 28}
\end{equation} 
depend only on $\bar \Omega$.
In figure 3 we show the dependence of $\epsilon_{2}$ (the solid black
curve) and $\lambda$ (the solid red curve) on $\bar
\Omega $. The solid green curve shows the dependence of the energy
transfer in the inertial frame, 
$\epsilon_{I}=\epsilon_{2}+\bar \Omega \lambda $. Note that the energy
transfer in the inertial frame is negative when $\bar \Omega > 2$.
The dotted curves show the dependencies of the corresponding
quantities calculated with only two global modes, the retrograde mode
with $\sigma_{1}=-1.063$ and the prograde mode with $\sigma_{2}=0.697$. 
Figure 3 shows  
that the dotted and the solid curves are very close to each
other. Therefore, only two modes are essential for our
problem.
The angular momentum transfer is equal to zero when 
$\bar \Omega=\bar \Omega_{crit}=1.6$, 
where $\epsilon_{crit} \approx 0.0065$. The condition $\Delta L_{2}=0$
may be easily realised in an astrophysical situation where the
inertial modes dominate the tidal response and the moment
of inertia of the star is sufficiently small. In this case 
the system quickly 
relaxes to the so-called state of pseudo synchronisation where 
$\Delta L_{2}=0$.  In figure 4 we show the dependence of 
$\Delta E/E_{*}$ on $\eta$ for this case.
The black curve
shows the contribution of the inertial waves, 
\begin{equation}
\Delta E_{2} \approx {0.0065\over (1+q)^{2} \eta^{6}}E_{*}. 
\label{eq 29} 
\end{equation}
The red curve
shows the contribution of the $f$ mode to the energy transfer
calculated for a non-rotating star (PT) and the green curve shows the
same quantity but calculated when $\bar \Omega =\bar \Omega_{crit}$
(IP)
\footnote{Note a misprint made in (IP). Their equations (60), (61) and (64)
must by multiplied by factor $\pi^2$. This 
has no influence on the conclusions of the paper.}. 
It is clear from this figure that when $\eta > 8$ 
the inertial waves dominate the tidal response. 
For a planet with 
Jupiter mass and radius circularising from large eccentricity
to attain a final period of $6$ days, $\eta \sim 18$ (IP), such that
inertial modes dominate by a wide margin.

\vspace{-0.7cm} 
\section{Conclusions}

In this Paper we formulate a new self-adjoint approach to the problem
of oscillations of uniformly rotating fully convective star and apply
this approach to the tidal excitation  
of the inertial waves and  tidal capture. The
approach can be used in any other problem 
where oscillations with typical frequencies of the order of the 
angular velocity are important. It can also be extended to the case of
convectively stable stars. The oscillations with higher
frequencies can also be treated in framework of an extension of our
formalism.

We show that the tidal response   due to inertial waves can be
represented as a spectral decomposition over normal modes.
This formalism is general enough to be applied even when
the spectrum is not regular and discrete.
For full polytropes,
only a few global modes with a large scale distribution 
of the perturbed quantities over the star contribute significantly
to the response. We find  general
expressions for the  energy and angular momentum transfer between the orbit 
and inertial waves. It is shown that when the inertial waves 
dominate the tidal response and a state of pseudo 
synchronisation is reached the energy transfer is determined by 
simple equation (\ref{eq 29}). As follows from that equation the 
energy transfer does not decrease exponentially with the orbital 
periastron and
therefore at large periastra the energy transfer due to inertial waves
dominates over the energy transfer due to excitation of the fundamental
mode. This can be very important for the problem of tidal
circularisation of the extra solar planets (see above,  IP and references
therein), and can be of interest for other problems, such as
eg. the problem of tidal capture of convective stars in globular
clusters. The application of our approach to  extra solar planets
will be discussed in a separate publication.

\vspace{-0.7cm}
  
\section*{Acknowledgements}

\vspace{-0.1cm}

PBI has been supported in part
by RFBR grant 04-02-17444.

\vspace{-0.4cm}

\bsp

\label{lastpage}

\end{document}